\documentclass[aps,onecolumn,longbibliography,superscriptaddress,doublespace,nofootinbib]{revtex4-2}
\usepackage{caption}
\usepackage{subcaption}

\usepackage{graphicx,color,appendix,physics,wasysym,xcolor,xparse}
\usepackage[margin=1.2in]{geometry}
\usepackage[english]{babel}
\usepackage{amsmath,amssymb,amsfonts,mathtools,mathdesign,mathstyle}
\usepackage{euscript,braket,natbib}
\definecolor{brickred}{rgb}{0.8, 0.25, 0.33}
\definecolor{celestialblue}{rgb}{0.29, 0.59, 0.82}
\definecolor{cornflowerblue}{rgb}{0.39, 0.58, 0.93}
\definecolor{denim}{rgb}{0.08, 0.38, 0.74}
\definecolor{armygreen}{rgb}{0.29, 0.33, 0.13}
\definecolor{cardinal}{rgb}{0.77, 0.12, 0.23}
\definecolor{carnelian}{rgb}{0.7, 0.11, 0.11}
\usepackage[colorlinks=true, urlcolor=black, citecolor=blue,linkcolor=red,citebordercolor={1 0 0},linkbordercolor={0 0 1}]{hyperref}

\usepackage{titlesec}
\usepackage{verbatim,multirow}

\usepackage{scrextend}

\usepackage[utf8]{inputenc}
\usepackage[T1]{fontenc}
\usepackage{lmodern}

\begin{document}
	\title{Quantum non-Markovianity: Overview and recent developments}
	\author{U. Shrikant}
	\email{shrikant.phys@gmail.com}
	\affiliation{Department of Physics, Indian Institute of Technology Madras, Chennai - 600036, India.}
	\author{Prabha Mandayam}
	\email{prabhamd@physics.iitm.ac.in}
	\affiliation{Department of Physics, Indian Institute of Technology Madras, Chennai - 600036, India.}
	\affiliation{Center for Quantum Information, Communication, and Computing, Indian Institute of Technology Madras, Chennai - 600036, India.}

	\begin{abstract}
		In the current era of noisy intermediate-scale quantum (NISQ) devices, research in the theory of open system dynamics has a crucial role to play. In particular, understanding and quantifying memory effects in quantum systems is critical to gain a better handle on the effects of noise in quantum devices. The main focus of this review is to address the fundamental question of defining and characterizing such memory effects -- broadly referred to as quantum non-Markovianity -- from various approaches. We first discuss the two-time-parameter maps approach to open system dynamics and review the various notions of quantum non-Markovianity that arise in this paradigm. We then discuss an alternate approach to quantum stochastic processes based on the quantum combs framework, which accounts for multi-time correlations. We discuss the interconnections and differences between these two paradigms, and conclude with a discussion on necessary and sufficient conditions for quantum non-Markovianity.
	\end{abstract}
	
	\maketitle

\section{Introduction}
A quantum system is said to be \textit{open} when it interacts with its environment \cite{breuer2002theory}. As such a system evolves, it builds up correlations, such as entanglement, with the environment \cite{vega2017dynamics}. This in turn results in decoherence and dissipation \cite{breuer2016colloquium}, which are known to be generally detrimental to quantum information tasks. The study of open system dynamics has thus become more important than ever, in today's era of noisy intermediate-scale quantum (NISQ) devices \cite{preskillNISQ}. Of particular interest is the characterization of memory effects, or, quantum non-Markovianity, that arises due to strong system-environment coupling. A precise and universal definition of non-Markovianity has remained elusive, and understanding its origins and characteristics is pertinent for emerging quantum technologies.

The study of open quantum system dynamics has been formalized in a number of approaches, starting with the traditional approaches in~\cite{breuer2002theory,banerjee2018open}, to operational characterizations~\cite{pollock2018non}, and more recently, approaches based on quantum collision models \cite{ciccarello2022quantum,campbell2021collision}. From the quantum information theory point of view, system dynamics are represented by \textit{quantum dynamical maps}, which are linear, completely positive (CP) and trace preserving (TP) maps. Such maps, referred to as \textit{quantum channels}, can be described using an operator-sum representation (the so-called Kraus representation)~\cite{nielsen_chuang_2010}, which can be derived by tracing out the environment degrees of freedom from the full system-environment unitary dynamics. 

Open system dynamics may be broadly classified as Markovian and non-Markovian. In the natural sciences, a process is said to be Markov if the future outcomes of the measurement on the system are independent of the past ones, conditioned on the present. When such past-future independence fails, or, when the environment retains the history of the system then the process is said to be non-Markovian. Over the past decade, there has been much effort focused towards characterizing, witnessing and quantifying non-Markovianity in the quantum domain. A number of witnesses and measures have been proposed, based on divisibility \cite{RHP10}, distinguishability (or trace distance) \cite{breuer2009measure}, fidelity \cite{rajagopal2010kraus}, quantum channel capacity \cite{bylicka2014non,pineda2016measuring}, accessible information \cite{fanchini2014non-Markovianity}, mutual information \cite{luo2012quantifying}, quantum discord \cite{alipour2012quantum}, interferometric power \cite{dhar2015characterizing} and deviation from semigroup structure \cite{wolf2008assessing,utagi2020temporal}, to name a few. 

However, a precise and universal definition of quantum (non-)Markovianity continues to remain one of the unsolved problems in open systems theory~\cite{li2018concepts}. The traditional approach to quantum non-Markovianity does not have a well-defined classical limit and lacks a clear operational interpretation \cite{pollock2018operational}. In fact, the traditional approach characterizes dynamical processes either via one-parameter semigroups of dynamical maps or two-parameter families of maps that are divisible and indivisible, thus incorporating only two-time correlations functions of the environment. The results emerging from such approaches do not necessarily generalize to situations where multi-time correlations become prominent. A new approach known as the process tensor formalism promises a solution to this problem through complete tomographic characterization of a quantum stochastic process by taking into account multi-time correlations as well as the (possibly unknown) initial system-environment (S-E) correlations \cite{modi2012operational}, offering an operationally motivated characterization of open systems which the previous approaches need not provide. 

The present review attempts to survey this active area of defining and characterizing quantum non-Markovian dynamics. While there have been a few good reviews on this topic in the literature in the past -- see, for example, \cite{RHP14,breuer2016colloquium,breuer2016nonmarkovian,vega2017dynamics,li2018concepts} -- our article focuses on the more notable recent developments aimed at detecting and quantifying non-Markovianity via temporal quantum correlations. After briefly reviewing the well-known definitions based on CP-divisibility \cite{RHP10} and distinguishability \cite{breuer2009measure} which are only necessary but not sufficient indicators of non-Markovianity, we discuss a measure proposed by \cite{chen2016quantifying} based on temporal steerable correlations and subsequently that proposed by \cite{shrikant2021causal} based on causality measure arising out of pseudo-density matrix. However, as we note in this review that the definitions based on quantum temporal correlations are only sufficient but not necessary indicators of non-Markovianity. Later, we discuss the recent approaches in which multi-time correlations are taken into account, specifically the process tensor framework proposed by \cite{pollock2018non,pollock2018operational} and a definition of non-Markovianity based on conditional past-future correlations proposed by \cite{budini2018quantum,budini2019conditional,budini2022non-operational}. 
Specifically, we address the issue of necessary and sufficient criteria for quantum non-Markovianity in this review. 
 
\textit{A note on terminology}: (i) We use \textit{system} to refer to an open quantum system (ii) \textit{environment} refers to a quantum environment having quantum degrees of freedom , unless otherwise stated; (iii) the \textit{master equation} is an equation that describes the reduced dynamics of the system alone, after tracing out the environment degrees of freedom; (iv) the word \textit{correlations} implies quantum correlations unless otherwise stated.

 The rest of this review is structured as follows. In Sections \ref{sec:part-1-master} and \ref{sec:part-1-maps}, we briefly review the well-known master equation and dynamical maps approaches to open system dynamics. In Section \ref{sec:measures-spatial} we discuss some of the famous measures of non-Markovianity including the ones based on distinguishability of states and completely positive (CP) divisibility. We also briefly note some of the measures that are based on quantum correlations. In Section~\ref{sec:measures-temporal}, we review some recent measures that are based on correlations in time, namely temporal steering and temporal non-separability, and note an important relation between the two. Interestingly, these measures are known not to be strictly equivalent, as we discuss in Section \ref{sec:failures}, leaving open the question of equivalence between the measures based on temporal steerable weight and that based on causality measure.
 
Sections \ref{sec:initial-correlationA} and \ref{sec:initial-correlationB} form an interlude where we discuss some curious features of open systems and system-environment (S-E) correlations, and mention some recent developments. We then move on to Part II in Section~\ref{sec:process-tensor}, where we mainly focus on the frameworks that overcome the limitations of the existing two-time maps. Given that a notion of Markovianity exists, namely the independence of future outcomes on past measurement results, non-Markovianity is commensurate with a notion of causality and causal influence of past history on the future of evolution. In Sec.\ref{sec:process-tensor} we present the notion of non-Markovianity based on an operational framework called process tensor and discuss various features, along with mentioning recent progress. In Sec.\ref{sec:CPF-correlations}, we review the notion of non-Markovianity based on conditional past-future independence which is operationally motivated and yet does overcome the limitation of previous approaches. In Section \ref{sec:experimental}, we briefly review some of the aspects of non-Markovian dynamics in experimental settings. We conclude in Section \ref{sec:conclusion}, with a brief discussion on necessary and sufficient criterion for a witness and measure of non-Markovianity for any arbitrary quantum stochastic dynamics and provide a note on future prospects.
 
\section{Part I: Two-time quantum dynamical maps and non-Markovianity \label{sec:part-1}}
\subsection{The master equation \label{sec:part-1-master}}
Traditionally, the reduced dynamics of a system coupled to an environment is described by a Nakajima-Zwanzig master equation, also called the time-nonlocal equation, of the form
\begin{equation}
    \dot{\rho}(t) = -\frac{i}{\hbar}[H_S,\rho(t)] + \int_{t_0}^{t} \mathcal{K}_{t,\tau}[\rho(\tau)] d\tau,
    \label{eq:nz1}
\end{equation}
$\forall t \ge \tau \ge 0$, where $\dot{\rho} = \frac{d\rho}{dt}$ and $H_S$ is the system Hamiltonian. The linear map $\mathcal{K}_{t,\tau}$ incorporates the non-Markovian memory effects that may be present in the system's evolution. One may go from the time-nonlocal equation to a time-local one by assuming that there exists a linear map $ \Phi $ which is invertible at all times, that is $\Phi \Phi^{-1} = I$, such that \cite{andersson2007finding},
\begin{align}
    \dot{\rho}(t) &= \int_{t_0}^{t} d\tau \, (\mathcal{K}_{\tau,t} \circ \Phi_\tau)[\rho(0)] \nonumber \\
    &= \int_{t_0}^{t} d\tau \, (\mathcal{K}_{\tau,t} \circ \Phi_\tau \circ \Phi^{-1}_t) \,\Phi_t[\rho(0)] = \mathcal{L}_t[\rho(t)].
\end{align}
$\mathcal{L}_t$ is the time-local \emph{generator} or the \emph{Lindbladian}, which is a linear super-operator on the space of density operators. When the corresponding dynamical map is non-invertible, it is not necessary that a time-local master equation should exist \cite{andersson2007finding}, although one may make use of Moore-Penrose pseudo-inverse \cite{RHP14} or generalized inverse and still be able to construct a generator at least for divisible dynamics \cite{chakraborty2021construction}. That is, existence of a master equation is sufficient to imply the existence of a corresponding dynamical map, however the converse is not true \cite{li2018concepts}. 

In order to arrive at an exact Lindbladian $\mathcal{L}$, one makes the famous Born-Markov approximation [see Sec. \ref{sec:part-1-maps}], under which the time-local evolution of the open system is described by the famous Gorini-Kossakowski-Sudarshan-Lindblad (GKSL) equation \cite{sudarshan_stochastic_1961,lindblad1976}, which in its canonical form reads,
\begin{align}
\Dot{\rho} = \mathcal{L}[\rho] = \sum_j \gamma_j \bigg(L_j \rho L_j^\dagger - \frac{1}{2}\{L_j^\dagger L_j, \rho\} \bigg).
\label{eq:gksl}
\end{align}
Here $\Dot{A} = \frac{dA}{dt}$ for any time-continuous operator $A$ and the linear operators $L_j$ are called the \textit{Lindblad operators} or simply the \textit{jump operators}. The dynamics described by Eq. (\ref{eq:gksl}) is called time-homogeneous Markovian. Generally, the decay rates $\gamma_j$ may be time-dependent. In this case, the Born-Markov approximation does not hold, but the rotating approximation is retained, so that the master equation modifies to the time-dependent GKSL-like equation (in the canonical form):
\begin{align}
\Dot{\rho} = \mathcal{L}(t)[\rho] = \sum_j \gamma_j(t) \bigg(L_j(t) \rho L_j^\dagger(t) - \frac{1}{2}\{L_j^\dagger(t) L_j(t), \rho\} \bigg) .
\label{eq:gksl-like}
\end{align}
Now the jump operators themselves are time-dependent along with the decay rates $\gamma_j(t)$. The process in Eq. (\ref{eq:gksl-like}) is termed time-inhomogeneous Markovian when all the decay rates $\gamma_j(t)$ are positive for all times. When at least one of the decay rates is negative for a certain interval of time-evolution, then the process is termed non-Markovian \cite{garraway1997nonperturbative,breuer2002theory}. 

The price one pays for going from a nonlocal to a local description is that the generator may become highly singular \cite{chruscinski2010non}, which makes the solution to dynamics analytically hard. Indeed, the nonlocal equation might be more natural and easy to handle in certain physical situations, see Ref. \cite{megier2020interplay,megier2020evolution}. The distinction between the notions of Markovianity in Eq. (\ref{eq:gksl}) and Eq. (\ref{eq:gksl-like}) becomes clearer when one looks at the properties of the corresponding dynamical maps, which we discuss next. 

\subsection{Quantum dynamical maps \label{sec:part-1-maps}} 
From a quantum information theoretic perspective, the general time evolution of a quantum system is described by a quantum dynamical map, which takes density operators to density operators. Since the environment is generally a many-body system with many degrees of freedom, it becomes difficult for an experimenter to fully control it. Therefore, studying the reduced dynamics of the system in a consistent manner becomes useful. 

\begin{widetext}
	\begin{figure}
		\includegraphics[width=0.8\textwidth]{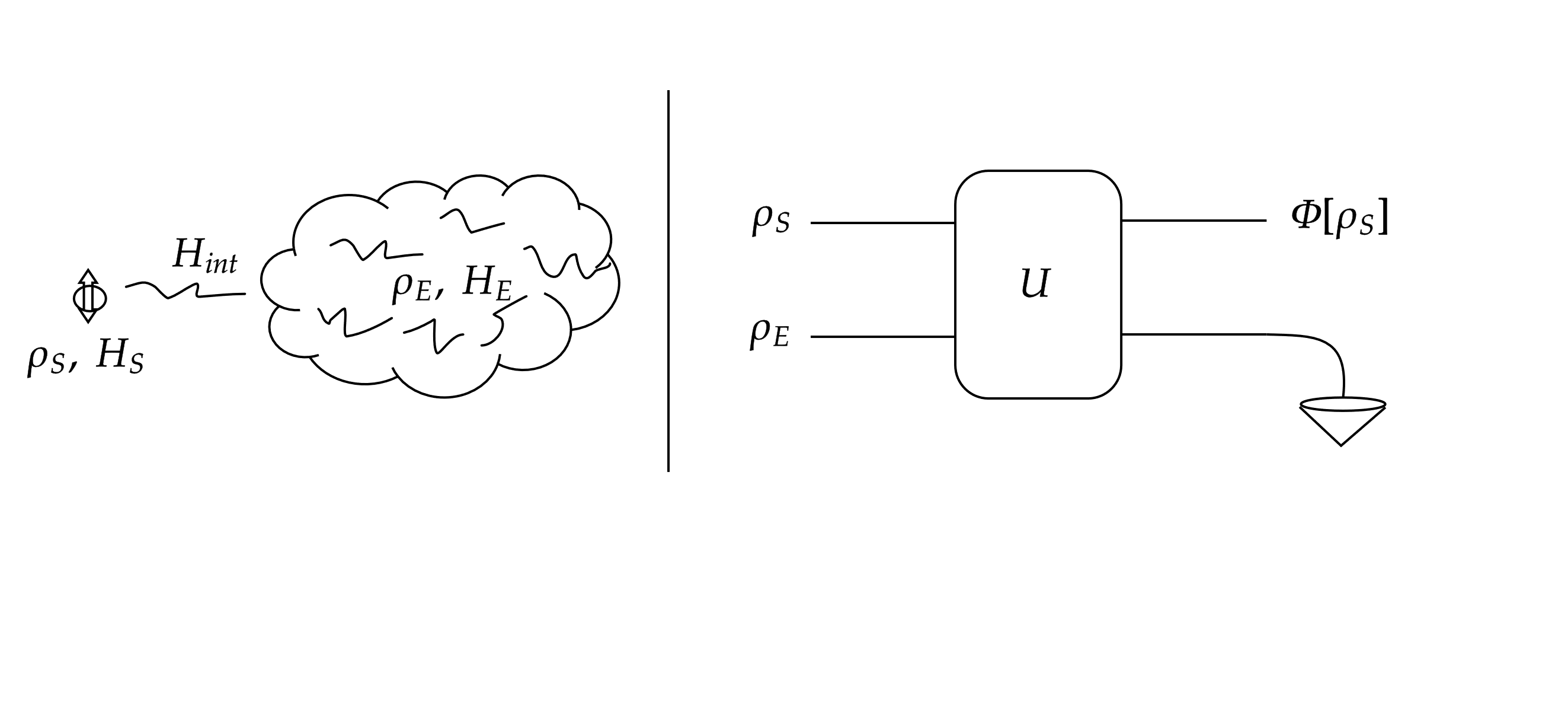}
		\caption{Left: A simple representation of a qubit interacting with environment degrees of freedom. Right: A simple model of quantum operation after tracing out the environment from the global unitary $U$. Here, $\rho_S$ and $\rho_E$ are system and environment states, respectively, and $H_S$, $H_E$, and $H_{int}$ are system, environment and S-E interaction Hamiltonians, respectively.}
		\label{fig:model}
	\end{figure}
\end{widetext}

Figure (\ref{fig:model}) depicts a simple example of an open quantum system, namely, qubit interacting with an environment. Let us denote the system Hamiltonian (also called the free Hamiltonian) as $H_S$ and the environment Hamiltonian as $H_E$. The interaction Hamiltonian $H_{int}$ determines the nature of system-environment interaction and the coupling with the environment. The total S-E evolution may be represented by a global unitary $U = \exp \{-\frac{i}{\hbar} (H_S + H_E + H_{int}) t\}$. The effect of the environment on the system is often called \emph{quantum noise} in the context of quantum information processing and the open-system evolution is termed noisy evolution in contrast to closed-system evolution which is unitary. 

Tracing out the environment degrees of freedom gives rise to the operator-sum representation of the effect noise on the system, which falls under the broad formalism of quantum operations. Mathematically, the effect of the environment of the system is represented by a set of linear operators on the system, called the Kraus operators, and the reduced dynamics of the system is obtained as follows.
\begin{align}
	\rho(t) &= \text{Tr}_E[U (\rho_S(0) \otimes \rho_E) U^\dagger ] = \sum_j \bra{e_j} U (\rho_S(0) \otimes \ket{e_0}\bra{e_0}) U^\dagger \ket{e_j},
	\label{eq:env-trace}
\end{align}
where the states $\{\ket{e_j}\}$ represent environment degrees of freedom. Equivalently, Eq.~\ref{eq:env-trace} can be written for any input system state $\rho$, in the so-called \emph{Kraus form}, as, \cite{sudarshan1961stochastic,kraus1971general,choi1975completely,kraus1983states}
\begin{align}
	\rho(t) = \Phi(t)[\rho]  &= \sum_j K_j(t) \rho K^\dagger _j (t).
	\label{eq:kraus}
\end{align}
Here, $K_j \equiv \bra{e_j}U \ket{e_0}$ are called the \textit{Kraus operators} which obey $\sum_j K^\dagger_j K_j \le I $. This operator-sum representation is an important and powerful tool today in the context of quantum information and computation \cite{nielsen_chuang_2010}. The map $\Phi$ in Eq.~\ref{eq:kraus} obeys the  time-homogeneous (or, time-independent) master equation (\ref{eq:gksl}), that is, $\Dot{\Phi} = \mathcal{L}[\Phi]$, whose solution is given by $\Phi = \exp \{t \mathcal{L}\}$, which is a one-parameter quantum dynamical semigroup. Similarly, a two-parameter map $\Phi(t,t_0)$ (or $\Phi(t)$ for simplicity setting $t_0 = 0$) obeys a master equation (\ref{eq:gksl-like}) of the form, $\Dot{\Phi}(t) = \mathcal{L}(t)[\Phi(t)]$, whose solution is given by
\begin{align}
  \Phi(t,t_0) = \mathcal{T} \exp \bigg\{ \int^{t}_{t_0}  \mathcal{L}(\tau) d\tau \bigg\}~, \label{eq:two-time-map}
\end{align}
where $\mathcal{T}$ is the time-ordering operator. In a sense, a map that is derivable from a given generator depends on the nature of $\mathcal{L}$. 

One must note that Eqs. (\ref{eq:gksl}), (\ref{eq:gksl-like}), and (\ref{eq:kraus}) are derived after assuming that the system-environment (S-E) state factors out at $t=0$, which need not be the case generally. Furthermore, the environment state $\rho_e$ is assumed to be \emph{fixed} for all times, in which case the Born-Markov approximation holds. Under the time-coarse-grained weak coupling limit, the evolution quickly ``forgets'' initial S-E correlations and tends to the Lindblad form \cite{royer1996reduced}. The existence of the time-independent Lindblad form \ref{eq:gksl} implies the following \textit{equivalent} statements: (i) The environment auto-correlation function is a delta function and corresponds to the white noise approximation. There is no back-action on the system due to static environment state, which also means that $\tau_E << \tau_S$, where $\tau_E$ is environmental correlation time and $\tau_S$ is the system relaxation time, in other words, the environment cannot store any information about the system's past evolution. This is the famous Born-Markov approximation; (ii) The system uniformly couples to all the degrees of freedom of the environment; (iii) The Lindbladian $\mathcal{L}$ is time-independent and the corresponding solution to the equation $\dot{\Phi}(t) = \mathcal{L}[\Phi(t)]$, is $\Phi(t) = \exp \{t \mathcal{L}\}$ which is a semigroup satisfying the property, $\Phi(t+\tau) = \Phi(t)\Phi(\tau)$, for all $0 \le \tau \le t$. 

Historically, any process that deviates from the semigroup structure was termed non-Markovian \cite{breuer2016colloquium}. Later developments in the quantum information community indicate that this is not the complete story, as we will elaborate in the following sections.

\subsection{Measures of non-Markovianity: Spatial domain \label{sec:measures-spatial}}

\subsubsection{CP-indivisibility} 

Divisibility is a property of dynamical maps which allows us to write a map as a concatenation of intermediate maps \cite{wolf2008dividing,wolf2008assessing,RHP10,chruscinski2011divisiblity,chruscinski2014degree,chruscinski2018divisibility,davalos2019divisibility}. In the classical case, a divisible (hence Markovian) stochastic process is given by concatenation of transition matrices. As far as the traditional approach is concerned, there is no known way of carrying the classical definition of non-Markovianity over to the quantum case. However, an approach based on divisibility states that a map $\Phi(t, 0) = \Phi(t, \tau)\Phi(\tau,0)$ is CP-indivisible if the intermediate map $\Phi(t,\tau)$ is not completely positive (NCP) in the sense that at least one of the eigenvalues of the matrix 
\begin{align}
    \chi = (\Phi(t,\tau) \times I) [\ketbra{\psi}{\psi}] \label{eq:choi-state}
\end{align}
is negative, where $\chi$ is called the Choi state \cite{choi1975completely} or Sudarshan B matrix \cite{sudarshan_stochastic_1961}, which is dual to the intermediate map $\Phi(t,\tau)$, and $\ket{\psi} = \frac{1}{\sqrt{2}}(\ket{00} + \ket{11})$ is a maximally entangled state. Based on the above considerations, the following measure of non-Markovianity was introduced by Rivas-Huelga-Plenio (RHP) in~\cite{RHP10}.
\begin{equation}
		\mathcal{N}_{\rm RHP}=\int_{0 ;\, \chi(t,\tau) < 0}^{t_{\rm max}} g(t) dt \, ;  \quad g(t)=\lim_{\tau\rightarrow 0^+}\frac{\Vert\chi(t,\tau)\Vert_1 -1}{\tau}, \label{eq:rhp}
	\end{equation}
	where, $\chi$ is the Choi matrix such that whenever $g(t) > 0$, the divisibility condition is broken and the time integral over the positive regions of $g(t)$ quantifies the quantum memory in the dynamics. Note that $\mathcal{N}_{\rm RHP}$ goes up to infinity, hence requires a suitable normalization so that it falls in the interval $\{0,1\}$, implying that for Markov processes $\mathcal{N}_{\rm RHP} = 0$.

In fact, for any finite $d$-dimensional open system, the RHP measure in Eq.~\ref{eq:rhp} is equivalent the one due to Hall-Cressor-Li-Anderson (HCLA) given by~\cite{hall2014canonical,shrikant2018non-Markovian},
	\begin{align}
	\mathcal{N}_{\rm RHP} = \frac{d}{2} \mathcal{N}_{\rm HCLA}; \quad	\mathcal{N}_{\rm HCLA} = - \int_{0; \; \gamma(t)<0}^{t_{\rm max}} \gamma(t) dt,
	\end{align}
where, the integration is carried over only the negative regions of the time-dependent decay rate $\gamma(t)$. Historically, it was understood that when maps generating the dynamics deviate from having a semigroup structure, one speaks of non-Markovianity \cite{breuer2016colloquium}. The condition $\gamma_j(t) \ge 0$ pertains to Markovian approximation for the time-dependent noisy dynamics, and the corresponding dynamical maps do not belong to a semigroup; such processes are termed \emph{time-dependent Markovian}. Note that the dynamics represented by the Lindblad form in Eq.(\ref{eq:gksl}) is strictly Markovian \cite{hall2014canonical}. When the decay rates in Eq.(\ref{eq:gksl-like}) are temporarily negative, the corresponding dynamical maps are no more CP-divisible.

\subsubsection{Information back-flow}

As an open system evolves, it generally sets up correlations with the environment and loses its information content irreversibly. However, that is true only when the system couples weakly to the environment. Under the strong coupling limit, the information might periodically return to the state from the environment, leading to \emph{information back-flow}. This also means that the environment remembers the history of evolution of the system. We briefly review here an approach by Breuer, Laine and Piilo (BLP)~\cite{breuer2009measure} to quantify this information back-flow based on the trace distance.

One may find a distance measure on the space of density operators that is contractive under the given CPTP map. Since matrix trace norm is known to be CP-contractive under a CPTP map \cite{nielsen_chuang_2010}, trace distance is one such natural candidate\footnote{In fact, Bures distance and quantum relative entropy are other measures that are contractive under CPTP maps and can witness non-Markovianity \cite{liu2013nonunital,megier2021entropic}.}. Mathematically, the trace distance is defined as,
\begin{align}
    \mathcal{D}(\rho_1,\rho_2) = \frac{1}{2} \Vert \rho_1-\rho_2 \Vert_1,
\end{align}
where $\Vert A \Vert_1 = \text{Tr}[\sqrt{A A^\dagger}]$ is the trace norm of an operator $A$. A map $\Phi(t_2,t_0)$, that takes a density operator from $t_0$ to $t_2$, is said to be Markovian if it satisfies the following data processing inequality.
\begin{align}
    \mathcal{D}\big(\Phi[\rho_1],\Phi[\rho_2]\big) \le \mathcal{D}\big(\rho_1,\rho_2\big),
    \label{eq:TD-monotone}
\end{align} for all times, where $\Phi[\rho]$ is given by Eq. (\ref{eq:kraus}). 

Breakdown of the monotonicity of trace distance shown in Eq.(\ref{eq:TD-monotone}) between any two orthogonal initial states under a CPTP map was used as a witness of non-Markovianity. The decrease in non-orthogonality of the states is interpreted as back-flow of information from the environment to the system. Note that when there is information back-flow, the intermediate map $\Phi(t_2,t_1)$ is not even positive, which in turn implies that the dynamical map $\Phi(t_2,t_0) = \Phi(t_2,t_1)\Phi(t_1,t_0)$ is positive (P-) indivisible~\cite{chruscinski2011divisiblity,chruscinski2014degree}. This is equivalent to saying that $\frac{d}{dt} \Vert \Phi[(\rho_1 - \rho_2)] \Vert_1 \ge 0$. Note that complete positivity of the map $\Phi(t_2,t_0)$ requires that the trace distance only decreases at $t=0$, but it can increase and decrease for $t >0$ due to P-indivisibility. 

Quantum non-Markovianity in the sense of P-indivisibility can be quantified as follows \cite{breuer2009measure}.
	\begin{align}
		\mathcal{N}_{\rm BLP} := \underset{\rho_1,\rho_2}{\rm max} \int_{{\frac{d\mathcal{D}}{dt}}>0}  dt \; \frac{d\mathcal{D}}{dt}~,
		\label{eq:cmeasureTD}
	\end{align}
where integration is done over positive slope of $\mathcal{D}$. Note that this criterion is only sufficient but not necessary, since it might fail as a witness of memory for some non-unital channels \cite{chruscinski2017detecting,liu2013nonunital} as well as for certain unital channels \cite{hall2014canonical}. In other words, P-indivisibility implies CP-indivisibility but the converse may not be true. 

Finally, we may note that as far as revival of quantum information and correlations is concerned, this is also possible when the environment is classical and therefore cannot store information about the system. In other words, for the revivals of information to take place, the environment need not be quantum. This observation calls for attention to re-evaluating the notion of system-environment back-action \cite{xu2013experimental}.

\subsubsection{Correlation-based measures}
We know that quantum mechanics allows for various forms of correlations namely, nonlocal correlations \cite{brunner2014bell} that violate Bell inequalities, steering \cite{uola2020quantum}, entanglement \cite{horodecki2009quantum}, entropic accord \cite{szasz2019accord} and quantum discord \cite{vedral2012discord}. While the RHP measure discussed in Eq.~\ref{eq:rhp} is based on entanglement, there exist various proposals based on different measures of correlation such as quantum discord \cite{alipour2012quantum}, mutual information \cite{luo2012quantifying} and accessible information \cite{fanchini2014non-Markovianity,haseli2014non-Markovianity,santis2019correlation}. Interestingly, some works show a peculiar relationship between non-Markovianity and certain forms of correlations, for example, quantum discord and non-Markovianity \cite{mazzola2011frozen,alipour2012quantum}. It was shown that a measure based on mutual information between the reduced system and an ancilla detects the range of non-Markovianity as BLP does \cite{luo2012quantifying}. Similar assertions may be made for any measure based on correlation between the system and an ancilla, for example the one proposed by \cite{RHP10}, in which entanglement is used to quantify non-Markovianity. However, it must be noted that some of them may be easier to calculate than the others. For instance, correlations between system-ancilla might be simpler compared to quantum discord between system and environment states which requires full knowledge of the S-E dynamics \cite{alipour2012quantum}.

As we have seen, a number measures and witnesses have been proposed based on correlations in space. However, recently, a few works have made use of \textit{correlations in time} to witness and measure non-Markovianity, which we take up in the next subsection.

\subsection{Measures of non-Markovianity: Temporal domain \label{sec:measures-temporal}}
As we noted before that the (spatial) correlations form a hierarchy, quantum temporal correlations also do so which was show by \cite{ku2018hierarchy} recently; that temporal nonlocality \cite{leggett-garg1985inequallity}, temporal steering \cite{chen2014temporal}, and temporal non-separability \cite{fitzsimons2015quantum} of the pseudo-density matrix (PDM) framework form the hierarchy. In the same paper, they also showed that temporal steering is a form of weak direct cause while temporal non-separability forms a stronger form of direct cause in quantum mechanics. Interestingly, temporal steering was quantified by temporal steerable weight (TSW) which was proven to be contractive under a divisible CPTP map, and was used to quantify quantum non-Markovianity by \cite{chen2016quantifying}. Here, we briefly review the measure based on TSW and causality measure based on PDM.

\subsubsection{Temporal steering}
Quantum steering is a way to prepare a part of an entangled bipartite state by making measurement on the other. In spatial steering, Alice performs positive operator valued measure (POVM) on the her system. Bob does not trust Alice and her apparatus either, and would wish to distinguish between the correlations established due to true manipulation of his local state and that of due to underlying classical local hidden variables. 
 
Similar to the steering in space with a given spatially entangled state, one may steer a state in time by making a measurement on the input state and sending it via a quantum channel followed by a complete quantum state tomography of the output state at the end of the channel. Now, we shall introduce the notion of temporal steerable weight. Alice performs a positive operator valued measure (POVM) measurement on an input state $\rho$ at $t=0$ transforming it into 
\begin{align}
    \rho_{a|x} = \frac{\Pi_{a|x} \rho \Pi_{a|x}^\dagger}{p(a|x)},
\end{align}
where $p(a|x) = \text{Tr}[\Pi_{a|x} \rho \Pi_{a|x}^\dagger]$ is the probability that an outcome $a$ occurs given that Alice preforms a measurement in the basis $x$. Now the state $\rho_{a|x}$ is sent to Bob down a noisy quantum channel $\Phi$ for a time $t$. When Bob receives the state at $t$ he performs a quantum state tomography to get the state $\sigma_{a|x}(t) = \Phi[\sigma_{a|x}(0)$. We may call the set of states $\sigma_{a|x}(t)$ as temporal assemblages, and let the unnormalized assemblage be $\sigma_{a|x}(t) \equiv  p(a|x) \sigma_{a|x}$. Now, by assumption, Bob doesn't trust Alice nor her devices, and he would want to distinguish the correlations due to Alice's measurements from the correlations that might have originated from a hidden variable $\lambda$, making the correlations to satisfy locality in time and realism. Therefore, we may represent the correlations that might be produced by such classical origins as 
\begin{align}
    \sigma^{US}_{a|x}(t) = \sum_\lambda P(\lambda) P(a |x, \lambda ) \sigma_\lambda, \label{eq:T-US}
\end{align}
where $\sigma^{US}_{a|x}(t)$ is the unsteerable assemblage and $P(a |x, \lambda )$ is the probability that an outcomes $a$ occurs given that Alice makes a measurement $x$, and $\lambda$ the hidden variable that might have influenced the outcome, in which case Bob will be able to write down his assemblage in the form (\ref{eq:T-US}), and when he can't, then he is sure that the state is prepared by Alice's measurement. Now, a measure of temporal steering was introduced by \cite{chen2016quantifying} called temporal steerable weight (TSW). In order to define TSW consider a convex mixture 
\begin{align}
    \sigma_{a|x}(t) = w \sigma^{US}_{a|x}(t) + (1 - w) \sigma^{S}_{a|x}(t) \quad \forall a,x.
\end{align}
Clearly, $\sigma_{a|x}(t)$ is an assemblage which might contain both unsteerable and steerable correlations, with the constraint $0 \le w \le 1$. The TSW for a given assemblage $\sigma_{a|x}(t)$ is defined by 
\begin{align}
    W^{\rm TS} = 1 - w',
\end{align} where $w'$ is the maximum value of $w$. TSW may be interpreted as the minimal steerable resources required to reproduce temporal steerable assemblage. That is, $W^{\rm TS}=0 \, \text{and} \, 1$ for minimal and maximal steerability, respectively. $w'$ may be obtained by semi-definite programming:
\begin{align}
  \text{Find} \quad  w' &= \text{max} \, \text{Tr} \sum_\lambda  w \sigma_\lambda, \nonumber \\
    \text{subject \, to} \quad \bigg(&\sigma_{a|x}(t) - \sum_\lambda q_\lambda(a|x) w \sigma_\lambda \bigg) \ge 0 \quad \forall a,x \nonumber \\
    & w \sigma_\lambda \ge 0 \quad \quad \forall \lambda,
\end{align}
where $q_\lambda(a|x)$ are the extremal values of $P_\lambda(a|x)$. 

Now, under the noisy quantum channel these correlations deteriorate and \cite{chen2016quantifying} have shown that $W^{\rm TS}$ is non-increasing under local operations. Therefore, we have the monotonicity condition
\begin{align}
    W^{\rm TS}_\rho \ge W^{\rm TS}_{\Phi[\rho]} \label{eq:TS-monotonicity}.
\end{align}
A Markov process satisfies the above condition, and a non-Markovian process violates it. Given this fact, a measure of non-Markovianity is nothing but the area under the positive slope of of $W^{\rm TS}_{\Phi[\rho]}$:
\begin{align}
    \mathcal{N}_{\rm TSW} = \int^{t}_{t=0 \,;\, \frac{d W^{\rm TS}}{dt} > 0 } \frac{d W^{\rm TS}}{dt} dt, 
\end{align}
which by the factor of $\frac{1}{2}$ is equivalent to 
\begin{align}
N := \int_{t_0}^{t_{\rm max}} \biggl| \frac{dW^{\rm TS}}{dt} \biggr|  dt + (W^{\rm TS}_{t_{\rm max}} -  W^{\rm TS}_{t_0} ).
\label{eq:TSmeasure2}
\end{align}

It is important to mention that $\mathcal{N}_{\rm TSW}$ is only a sufficient and not a necessary condition for non-Markovianity of $\Phi$. There may be channels that will be detected as Markovian by this measure while other measures may detect them as non-Markovian. Breakdown of monotonicity of TSW may be interpreted as information back-flow from the environment to the system, hence this measure captures the range of memory effects that BLP does. 

\subsubsection{Pseudo-density matrix}

Recently, attempts have been made to define \textit{states across time}, similar to the states defined in space \cite{zhang2021quantum,zhang2020quantum,cotler2018superdensity}. It was shown that both these states have different structure, and a construction by \cite{fitzsimons2015quantum,pisarczyk2019causal} called the pseudo-density matrix (PDM) - which is a state correlated in spacetime, allowing for a treatment of correlations in space and time on an equal footing. However, one should note that the framework is ambiguous for the systems of dimension other than prime power \cite{horsman2017can}.

Recently, \cite{shrikant2021causal} has defined a measure for non-Markovianity based on temporal correlation in PDM. Utilizing the fact that the most general PDM is constructed by making measurement before and after a system passes through a quantum channel, one could quantify quantum non-Markovianity in the quantum channel in a straightforward way. Here, we consider a qubit across time evolved through a quantum channel, for simplicity. Let $\rho$ be the input state and $\Phi(t_B,t_A)$ a quantum channel that takes a density operator $\rho_A$ on $\mathcal{L}(\mathcal{H}_A)$ at time $t_A$ to an operator $\rho_B$ on $\mathcal{L}(\mathcal{H}_B)$ at time $t_B$.  Then a two-point PDM is given by
\begin{align}
R_{AB} = \left(I \otimes \Phi \right)\left[\bigg\{\rho \otimes \frac{I}{2}, \mathsf{swap} \bigg\}\right]
\label{eq:pdm2}
\end{align}
where $\mathsf{swap} := \frac{1}{2} \sum_{i=0}^{3} \sigma_i \otimes \sigma_i$ and $\{\hat{a},\hat{b}\} = \hat{a}\hat{b} + \hat{b}\hat{a}$ is the anti-commutator of operators $\hat{a}$ and $\hat{b}$, and $\sigma_i$ are Pauli-X,Y and Z operators with $\sigma_0=I$. In fact, it can be shown \cite{horsman2017can} that the two-point PDM can be written as
\begin{align}
  R_{AB}  = \frac{1}{2} \bigg( \rho_A \otimes \frac{I_B}{2} \chi_{AB} + \chi_{AB} \rho_A \otimes \frac{I_B}{2} \bigg)
\end{align}
where $\chi_{AB}$ is the Choi state of the channel $\Phi$, which derives from the so-called start product. One must note that the PDM is hermitian and has unit trace, but is not positive semi-definite when it is constructed out of measurements made in time. The reason is that this framework considers the tensor product over the \textit{same} Hilbert space of the input and output density operators, in order to define a state across time. However, under partial trace it describes a positive semi-definite operator at each instant of time, which is consistent with the current formulation of quantum mechanics.

A measure of temporal correlations in PDM has been defined by \cite{pisarczyk2019causal}:
\begin{align}
F= \log_2\|R_{\rm AB}\|_1,
\label{eq:cmeasure}
\end{align}
 which implies that when $F > 0$ the state $R_{\rm AB}$ is temporally correlated. Since $F$ is non-increasing under local quantum operations, for a Markovian channel $\Phi$, the following condition holds:
\begin{align}
F_{\Phi[\rho]}(t) \ge F_{\Phi[\rho]}(t+\tau),
\label{eq:monotonicity}
\end{align}
with $t+\tau \ge t$. A non-Markovian (or CP-indivisible) channel breaks the monotonicity condition (\ref{eq:monotonicity}). Following \cite{RHP10,chen2016quantifying}, a measure was proposed by \cite{shrikant2021causal} as the area under the positive slope of $F_{\Phi[\rho](t)}$:
\begin{align}
\mathcal{N}_{\rm \small causality} := \underset{\rho}{\rm max} \int_{\sigma_{(\rho,\mathcal{E},t)}>0}  dt \; \sigma_{(\rho,\mathcal{E},t)}
\label{eq:cmeasure0}
\end{align}
where $$\sigma_{(\rho,\mathcal{E},t)}= \frac{dF}{dt}.$$ The above definition, by a factor of $\frac{1}{2}$, is equivalent to 
\begin{align}
\mathcal{N}_{\rm \small causality} := \underset{\rho}{\rm max} \int_{t_0}^{t_{\rm max}} \biggl| \frac{dF}{dt} \biggr|  dt + (F_{t_{\rm max}} -  F_{t_0} ).
\label{eq:cNMmeasure}
\end{align}
The integral (\ref{eq:cNMmeasure}) is such that for a non-Markovian process, the derivative of $F$ is positive and $\mathcal{N}_{\rm \small causality} > 0$. And for a time-dependent Markovian (or, CP-divisible) process, the derivative of $F$ is negative and hence $\mathcal{N}_{\rm \small causality}=0$. It has been shown, using a phenomenological process, that $\mathcal{N}_{\rm \small causality} > 0$ corresponds to negativity of decay rate in the master equation, which is equivalent to RHP definition. It was also shown that $\mathcal{N}_{\rm \small causality} > 0$ also corresponds to information back-flow, and hence understood to be equivalent to the BLP definition for a pair of states optimal for the process under consideration. This equivalence is due to the fact the process considered in \cite{shrikant2021causal} has single jump operator in the Lindbladian. Interestingly, however, \cite{shrikant2021causal} showed that this equivalence of PDM based measure and BLP measure breaks down when non-Markovianity solely originates from the non-unital part of the channel.

Here again, we mention that PDM based non-Markovianity measure is sufficient but not a necessary criterion for non-Markoviantiy. However, \cite{ku2018hierarchy} have shown that PDM correlations contain stronger form of quantum direct cause and temporal steerable correlations that of weaker form. There certain advantages of using PDM based measure over that on TSW. The PDM based measure (\ref{eq:cNMmeasure}) doesn't require any optimization procedure and it is easy to compute.  However, both measures do not require optimization over the input states as BLP requires, making these measures relatively easy to compute. A limitation of PDM is that, in current form, it is ambiguously defined for system with dimension other than prime power \cite{horsman2017can}. The full validity of these two measures requires further study.

\subsection{Equivalence of the measures and regimes of failure \label{sec:failures}}
In fact, there exists hierarchy of divisibility of map \cite{chruscinski2011divisiblity,chruscinski2014degree}, in which if a process is non-Markovian according to BLP measure, then the corresponding map $\Phi(t,0)$ is termed positive (P-)indivisible when the intermediate map $\Phi(t,s)$ is not even positive in the sense that it takes a positive state to a negative state. Whereas, a CP-indivisible map can be P-divisible. This prompts us that these definitions need not be equivalent, which is the case for an `eternally' non-Markovian Pauli channel \cite{hall2014canonical} for example which is CP-indivisible, but P-divisible. However, for certain non-unital channels BLP measure fails, and may require some modification \cite{liu2013nonunital}. BLP indicator essentially fails when the environment evolves \textit{independent} of the system \cite{budini2018maximally}.  It is interesting to note that when there is only a single decoherence channel, which corresponds to single jump operator in the Lindbladian, both CP- and P-indivisibility definitions coincide \cite{breuer2016colloquium} for any non-Markovian process. P- and CP-divisibility based witnesses, in general, coincide for bijective maps \cite{bylicka2017constructive}. Interestingly, \cite{chakraborty2019information} show that information back-flow and CP-indivisibility are equivalent notions for any open \textit{qubit} evolution. Recently, it has been noted that the negativity of decay rate is not sufficient to capture CP-indivisibility for non-invertible maps \cite{chruscinski2018divisibility}, particularly when there are multiple time-dependent decay rates in the master equation. Interestingly, P- and CP-divisibility as notions of Markovianity coincide for multi-level amplitude damping processes \cite{chruscinski2022markovianity}.

It must be noted that PDM contains correlations that characterize a form of strong direct quantum cause and temporal steerable correlations form that of a weaker form. It was noted in \cite{chen2016quantifying} that the measure based on TWS is necessary but sufficient to detect non-Markovianity. Therefore, it remains an open question if these measures for non-Markovianity vary in their ability of detecting weaker and stronger forms of non-Markovianity as RHP and BLP measures respectively do. So far, it is clear that the causality based measure by \cite{shrikant2021causal} is sufficient, but it is not yet known if it is also necessary as a non-Markovianity indicator.

However, these definition and measures, respectively detect and quantify non-Markovianity of only CP- and P-indivisible processes. Whereas it is known that there exist non-Markovian processes with colored environmental memory spectrum, hence non-Markovian \cite{yu2010entanglement,kumar2017nonmarkovian}, but are CP-divisible. These processes, even though, CP-divisible, can delay entanglement sudden death because of the quantum memory effect. It was also noted that when the generator of the dynamics depends on the initial time, it leads to a kind of memory effect in the dynamics on the level of master equation even when the dynamics is CP-divisible \cite{chruscinski2010non,benatti2012non,utagi2020temporal}. 

An interesting notion of memorylessness (or, Markovianity) was proposed by \cite{utagi2020temporal}, in which a dynamical map is said to Markovian (more precisely a semigroup), if the dynamical map is independent of the initial time. This notion was termed `temporal self-similarity' in the sense that the form of the map remains same throughout the dynamics. This notion is, in fact, commensurate with time-homogeneity of semigroup evolution. The motivation behind this notion was to find a witness and measure of non-Markovianity of certain kinds of noise such Ornstein-Uhlenbeck and Power-Law noise that have colored memory spectrum \cite{kumar2017nonmarkovian}, but yet are CP-divisible and hence oblivious to RHP measure. The measure based on the deviation from semigroup proposed by \cite{utagi2020temporal} is as follows. 

From Eq. (\ref{eq:two-time-map}) one may get the infinitesimal map 
\begin{equation}
    (\delta\Phi)\rho(t) = \mathcal{T}\exp\left(\int_t^{t+dt}\mathcal{L}(\tau)d\tau\right)\rho(t)=  (1+\mathcal{L}(t)dt)\rho(t).
    \label{eq:inf-map}
\end{equation}
From CJ isomorphism, the Choi state $\chi_{\small \Phi(t)}$ of the infinitesimal map (\ref{eq:inf-map}) is found to be $(d\ket{\psi^+}\bra{\psi^+} + \chi_{\small \mathcal{L}(t)}dt)$, where $\chi_A$ denotes the Choi state of the operator $A$ in question and $\ket{\psi^+} \equiv d^{-1/2}\sum_i \ket{i,i}$ is maximally entangled $d$-dimensional state. Some simple algebra gives one the Choi state of the generator. The authors use the time-averaged distance between the Choi states of the time-independent $(\mathcal{L})$ and time-dependent $(\mathcal{L}(t))$ generators to quantify non-Markovianity, given as,
\begin{equation}
\mathcal{N}_{\rm \small SSS} = \min_{\gamma} \frac{1}{T} \int_0^T \Vert \Delta L \Vert_1 dt ,
\label{eq:sss}
\end{equation}
where $T$ represents some time interval. Here, $\Delta L \equiv  \delta \chi_{\small \Phi(t)} - \delta\chi_{\small \Phi} = (\chi_{\small \mathcal{L}(t)} -\chi_{\small \mathcal{L}} )dt$. The minimization over time-independent $\mathcal{L}$ leads to the minimization  over all possible time-independent decay rates $\gamma$. Positive $\mathcal{N}_{\rm \small SSS}$ means that a process is non-Markovian even when it is CP-divisible. This feature makes this measure a sufficient and necessary criterion for non-Markovianity.

Recently, \cite{budini2018quantum,budini2019conditional} proposed a definition of non-Markovianity that can detect memory present even in CP-divisible processes \cite{silva2020detection}. Here, memory effect is associated with the breakdown of conditional past-future (CPF) independence (or to the existence of CPF correlations), which is calculated using only three (sufficient) consecutive measurements on the system and post-selection on the outcomes. Although past-future independence, shown by \cite{li2018concepts} to be equivalent to `composability', thence to the semigroup structure of the dynamical maps, one may expect that semigroup dynamics generates statistics that obey CPF independence of \cite{budini2018quantum,yu2019experimentalCPF}. 

From this section one understands that any witness that detects non-Markovianity of a CP-divisible process is necessary and sufficient. We discuss this in the Afterward through Part II. 

\section{Interlude: The problem of (initial) system-environment correlations \label{sec:S-E-CPness}}
As noted before, a master equation, under Born-Markov approximation, need not exist if the initial S-E correlations are taken into account. Before going into those details, it is pertinent to ask when does actually decoherence begin, given that the underlying system evolution is described by a CP map (i.e., assuming no initial S-E correlations). We briefly point out the literature in the next sub-section and then move on to the issues surrounding the physically viable description of quantum stochastic processes without needing the initial S-E state to be separable.

\subsection{S-E correlations, decoherence, and non-Markovianity \label{sec:initial-correlationA}} 
It is generally understood that decoherence takes place when the system degrees of freedom ``entangle'' with that of the environment; hence entanglement must be necessary for decoherence \cite{schlosshauer2007decoherence}. However, this common wisdom might be mistaken, as \cite{pernice2011decoherence} show that this holds only when the system state starts out as a pure state. If the system's initial state is a mixed state, then decoherence may begin well before the system and environment get entangled. This also suggests that classical correlations might suffice for decoherence to take place.

Given that correlations, classical or quantum, are responsible for the onset of decoherence, it is interesting to explore the relationship between S-E correlations and non-Markovianity. The earliest notion of quantum non-Markovianity actually goes back to the deviation from semigroup structure which arises out of the so called Born-Markov (BM) approximation \cite{breuer2016colloquium}, which basically means that system and environment are \textit{weakly} coupled for all times.

The connection of S-E correlations with non-Markovianity has attracted the attention of the quantum information community \cite{devi2011open,breuer2016colloquium,vega2017dynamics,li2018concepts}. The S-E joint state may start out as product state, and later due to strong coupling between the system and environment, there may be certain points in time the when S-E correlations either weaken or even break momentarily causing the open system dynamics to transition from being Markovian to non-Markovian. It was noted that the S-E correlations decrease when there is information back-flow from the environment to the system \cite{mazzola2012dynamical}. However, it was also shown by \cite{pernice2012system} that there need not be any relationship between the decrease in S-E quantum correlations (specifically S-E entanglement) and non-Markovianity. Interestingly, if the environment is classical, there may be maximally non-Markovian evolution without S-E back-action or information flow \cite{budini2018maximally}. When a two qubits are independently interacting with a classical random field, there may be revivals of classical, quantum discord and entanglement between them even when there is no back-action from the environment to qubits \cite{franco2012revival}.

\subsection{Initial S-E quantum correlations, CP evolution, and non-Markovianity \label{sec:initial-correlationB}}
As we have noted, system-environment (S-E) correlations play the central role in open systems. To describe the reduced dynamics of the system via Lindblad equation (\ref{eq:gksl}), the joint S-E state is assumed to be \textit{factorized}, i.e., $\rho_s(0) \otimes \rho_e$, at the initial time and the environment state $\rho_E$ is assumed to be fixed for all times. Now, given that initial S-E state in not a product, the reduced dynamics of the system is argued to be not-CP; see Ref. \cite{pechukas1994reduced,alicki1995comment,pechukas1995pechukas,shaji2005s,rodriguez2008completely,schmid2019why} for some early results. However, it is possible to have physically meaningful not-CP maps, given that the domain of validity is known where such a map would still output a positive state \cite{jordan2004dynamics}. Moreover, Pechukas's assignment map can be made linear by sacrificing either positivity or reasonable consistency \cite{rodriguez2010linear}. There also have been arguments for and against vanishing quantum discord as being necessary and sufficient condition for complete positivity \cite{shabani2009vanishing, brodutch2013vanishing,sabapathy2013quantum}. However, it is clearly established that when initial S-E correlations are classical the reduced dynamics can always be described by a CP map \cite{rodriguez2008completely}. 

\cite{brodutch2014complete} argued that, if there are no anomalous information back-flows from the environment to system, it is necessary and sufficient to describe the reduced dynamics of the system by a CPTP map.  Interestingly, it was shown earlier that a witness for initial S-E correlations upper bounds the witness of non-Markovianity based on BLP (or, information back-flow) criterion \cite{rodriguez2012unification}. This prompts further investigations into the problem of initial S-E correlations and non-Markovianity. Recently, \cite{strasberg2018response} introduced a measure which quantifies non-Markovianity even when the initial S-E state is entangled. A work by \cite{schmid2019why} argues why initial system-environment correlations do not imply the failure of complete positivity, from the point of view of quantum causality. \cite{ringbauer2015characterizing} proposed a method to characterize a superchannel by making measurements on the system alone even when it is correlated with the environment. A more general result by \cite{silva2019dynamics} says that it is still possible to have a $d^2$ (or less) number (i.e., a family) of CP maps describing $d$-dimensional system evolution with initial S-E (quantum) correlations, i.e., by doing local operations on the system, one could derive a set of CP maps that describe the S-E evolution with initially correlated state. 

This is still an active area of research where there is no clear consensus about the role of initial S-E (quantum) correlations in open system dynamics where complete positivity is paramount. \cite{alipour2022correlation} have put forth a technique that makes use of correlation parent operator that allows one to write a down a master equation with initial correlation within the weak coupling regime. A technique based on adapted projection operators was introduced by \cite{trevisan2021adapted}, in which they apply a perturbative method to model a global S-E evolution which incorporates fully general initial correlations. For other recent attempts that have been made to accommodate initial S-E correlations into a valid theory of open systems without sacrificing linearity and complete positivity, see Ref.~\cite{silva2019dynamics} and \cite{pollock2018operational,pollock2018non}, with specific aims of characterizing open system evolution operationally and allowing a quantum stochastic process to have an appropriate classical limit. We take this up in Section \ref{sec:process-tensor}.

\section{Part II: Multi-time correlations and non-Markovian processes \label{sec:process-tensor}}
So far, we have been considering only the two-time parameter dynamical maps that are related to two-time correlation functions of the environment. In fact, a quantum regression formula can be obtained through two-time maps which helps us relate its satisfaction to the semigroup evolution of the open system under the initial factorized state assumption \cite{li2018concepts}. However, recent interest has grown in reconsidering quantum multi-time processes \cite{lindblad1979non} generalizing the so-called quantum stochastic process \cite{sudarshan_stochastic_1961}, however in the light of operational quantum theory.

\subsection{Quantum regression}
Quantum regression hypothesis (QRH) or quantum regression formula (QRF) must be invoked for calculation of \textit{multi-time correlation functions}, without the knowledge of environmental degrees of freedom, that is only with the mean values of the operator on the system Hilbert space alone. However, while non-Markovian evolution \textit{must} violate QRH, Markovian evolution (in the sense of CP-divisibility) \textit{can} as well \cite{guarnieri2014quantum}. Under the weak coupling and singular coupling limit, semigroup dynamics obey QRH \cite{swain1981master,dumcke1983convergence,davies1974markovian,davies1976markovian}. It was shown that Born-Markov (BM) approximation implies no-back-action \cite{swain1981master}, which means that the environment doesn't evolve \textit{due to} the interaction with the system, and the S-E state remains factorized for all times. It has been shown that CP-divisible dynamics violate QRH \cite{guarnieri2014quantum}. This prompts us that even RHP or CP-divisibility based criterion of non-Markovianity, like BLP criterion, is also not necessary but sufficient. Therefore, one is tempted to conjecture that violation of QRH is necessary and sufficient condition for a quantum stochastic process to be non-Markovian.  

\subsection{Process tensor}
A classical stochastic process $(X,t)$ is a collection of joint probability distribution of a system's state: 
\begin{align}
    P(X_j,t_j ; X_{j-1},t_{j-1} ; \cdots ; X_1,t_1 ; X_0,t_0 )\;\; \forall j \in N,
\end{align}
which must satisfy Kolmogorov consistency conditions, where $X$ is the random variable defining the process and $t_j$ are the time instances at which $X_j$ outcomes occur with probabilities $P(X_j,t_j)$. Then, a Markov process or chain satisfies the following condition:
\begin{align}
    P(X_j,t_j | X_{j-1},t_{j-1} ; \cdots ; X_1,t_1 ; X_0,t_0 ) = P(X_j,t_j | X_{j-1},t_{j-1} )\quad  \forall j \in N,
\end{align}
where $P(A|B)$ denotes probability of obtaining A given B. 

It is not straightforward to define similar joint probability distribution in the quantum domain. What is the most general way one may represent a physical process which also has an operational meaning? Quantum combs formalism is one such method
\cite{chiribella2009theoretical,pollock2018operational}. As opposed to the traditional approach discussed in Part I of this review, where only two-time correlations are considered to define quantum non-Markovianity, process tensor formalism defines non-Markovianity based on the presence of temporal correlations in a multi-time quantum stochastic process \cite{pollock2018operational,pollock2018non}. Such descriptions of open systems might have specific implications for designing information processing tasks in the laboratory, where two-time correlations might not capture the full characteristics of an underlying process. 

\begin{figure}
	\centering
	\includegraphics[width=0.62\textwidth]{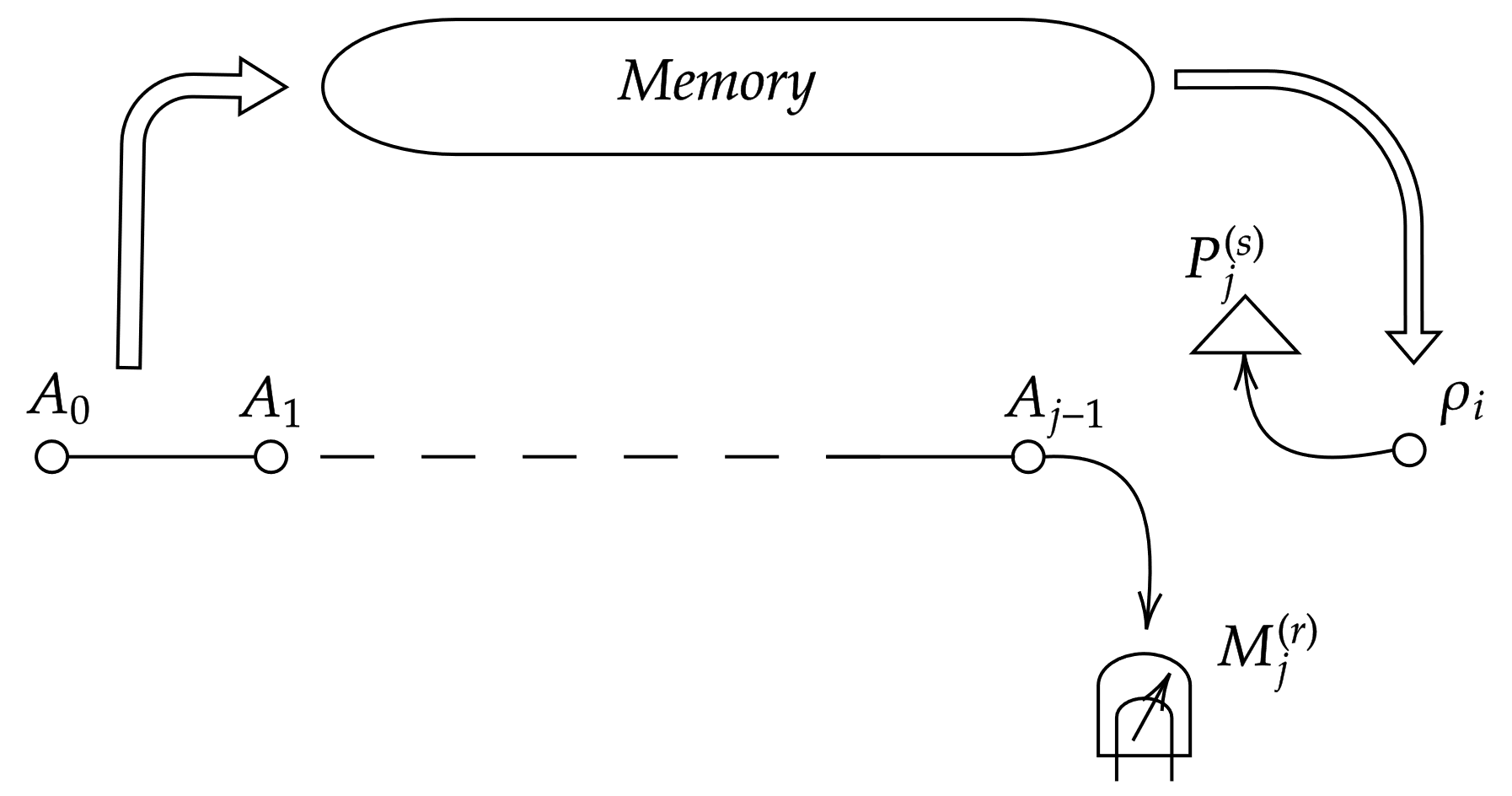}
	\caption{Schematic representation of process tensor framework with memory. $A_j$ are the control operations and $M^{(r)}_j$ and $P^{(s)}_j$ are the measurements and re-prepared states, respectively.} 
 \label{fig:process-tensor}
\end{figure}

 A quantum process is characterized by $j$ steps, with $0 \le j \le N$, when the system's state can be predicted at any instant $j$. The system is subject to intermediary operations $A$ which may be interrogations, manipulations, unitaries, or CP maps in general, and let $\{A_j\}$ and $\{M_j$\} be the set of local operations and measurements, respectively, on the system. `Process Tensor' $T_{j:0}$ is a mapping from the sequence of operations [see Figure (\ref{fig:process-tensor})]
\begin{align}
    \mathbf{A}_{j-1:0} := \{A_{j-1};A_{j-1};\cdots A_1 ;A_0\}
\end{align}
to the state $\rho_j$
\begin{align}
    \rho_j = T_{j:0}[\mathbf{A}_{j-1:0}]. \label{eq:PT-definition}
\end{align}
In general, $T_{j:0}$ satisfies (i) linearity: $T[a\mathbf{A}+b\mathbf{B} = a T[\mathbf{A}]+b T[\mathbf{B}]$; (ii) complete positivity, that it is a positive map on an extended space: $(T \otimes I)[\mathbf{A}_{SA}] = \rho_{SA} \ge 0$; (iii) containment: If $j \ge j' \ge k' \ge k$, then $T_{j':k'}$ is contained in $T_{j:k}$, in the sense that a process tensor on fewer times is not obtained by summing over excessive times, rather by appending identity maps for the excessive times.

Process tensor can be used to describe open quantum system dynamics. Let $\mathcal{U}_{j:j-1}$ be the S-E unitary that acts on S-E space as $\mathcal{U}_{k:l}[\rho_l^{SE}] = U_{k:l} \rho_l^{SE} U^\dagger_{k:l} = \rho_k^{SE}$, with $U_{k:l} U^\dagger_{k:l} = I$, and the system state be given by tracing over the environment $\rho_j = {\rm Tr}_E[\rho_j^{SE}]$, then the total dynamics follows:
\begin{align}
    \rho_j^{SE} = \mathcal{U}_{j:j-1}A_{j-1}\mathcal{U}_{j-1:j-2} \cdots A_1 \mathcal{U}_{1:0}A_0 [\rho_0^{SE}],
\end{align}
where $\rho_0^{SE}$ is the initial S-E state. Note that $T_{j:0}$ itself can be given a Kraus decomposition \cite{pollock2018non}. In an appropriate limit, the process tensor reduces to the conventional two-time maps picture of open system evolution:
\begin{align}
    \rho_j &= {\rm Tr}_E \big( U_{j:0} \rho_0^S \otimes \rho_0^E U^\dagger_{j:0} \big) = \Phi_{j:0}[\rho_0],
\end{align}
where $\Phi_{j:0} $ is a CPTP map. Therefore, $\rho_j$ can be obtained from the process tensor by applying identity as intermediate control operations on the system:
\begin{align}
    \rho_j = T_{j:0}[I; I ; \cdots I ; A_0].
\end{align}
The advantage of using process tensor framework is that one may map temporal correlations in the process to a many-body entangled state. A $j$-step process can be mapped to a many-body state via generalized CJ-isomorphism:
\begin{align}
    T_{j:0}[\mathbf{A}_{j-1:0}] = {\rm Tr}_S [\xi_{j:0} (I_S \otimes A_{j-1} \otimes I \otimes \cdots  \otimes A_0 \otimes I [(\Psi^+)^{\otimes j-1}])],
\end{align}
where the partial trace is over all subsystems except the one corresponding to the output of the $T_{j:0}$, and $\Psi^+$ is a maximally entangled bipartite density operator. In other words, the action of process tensor $T_{j:0}$ on the sequence of operations $\mathbf{A}_{j-1:0}$ is equivalent to projecting the Choi state $\xi_{j:0}$ onto the Choi state of $\mathbf{A}_{j-1:0}$. Here, $\xi_{j:0}$ is called the generalized Choi state of the $j$-step process that is mapped to a $(2j+1)$-body state which has a matrix-product-operator representation \cite{peres-garcia2007matrix}, and the Choi state $\xi_{j:0}$ has the bond dimension that is bounded by the effective dimension of the environment \cite{pollock2018non}.  

Given the above framework, we are in a position to discuss a definition of quantum (non-)Markovianity from an operational point of view. Let us denote the system at time step $i$ as a function of control operations: $\rho_i = \rho_i (\mathbf{A}_{i-1:0})$.  After the measurement, the system is re-prepared in a state $P_j^{(s)}$, selected randomly out of a set $\{ P_j^{(s)} \}$. The procedure of measuring and re-preparing the system introduces a `causal break' between the past $k \le j$ and future $i > j$ [See Fig.(\ref{fig:process-tensor})]. Similar to the classical definition, the Markov condition in the quantum regime reads
\begin{align} 
\rho_i \big(P_j^{(s)} | M_j^{(r)} ; \mathbf{A}_{j-1:0} \big) = \rho_i \big(P_j^{(s)} \big) \quad \forall \{ P_j^{(s)}, M_j^{(r)},\mathbf{A}_{j-1:0} \} \quad \text{and} \quad \forall \; i, j \in \{0,N\} . 
\end{align}
On the contrary, a quantum process is non-Markovian iff there exist at least two distinct, independent operation sets $\{ M_j^{(r)} ; \mathbf{A}_{j-1:0} \}$ and $\{ M^{\prime (r^{\prime})}_j  ; \mathbf{A}_{j-1:0}^\prime \}$, such that the resulting two conditional states are different:
\begin{align}
    \rho_i \big(P_j^{(s)} | M_j^{(r)} ; \mathbf{A}_{j-1:0} \big) \neq \rho_i \big(P_j^{(s)} | M^{\prime (r^{\prime})}_j   ; \mathbf{A}_{j-1:0}^\prime \big) .
\end{align}

The system itself cannot carry the information into the future across the causal breaks. An environment and the S-E correlations carry the information about the initial state of the system across causal breaks [See Figure (\ref{fig:process-tensor})], and this is what is called quantum non-Markovian memory in the process tensor. In order to quantify the memory in the process, one makes use of the mapping from temporal correlated process tensor to a many-body state via generalized CJ-isomorphism. Given the intermediate maps $\Phi_{j:j-1}$ that take a state from time step $j-1$ to $j$, a Markov process is fully characterized by its Choi state on the tensor product of initial system state and Choi states of independent CPTP two-time maps:
\begin{align}
    \xi^{\rm \small Markov}_{j:0} &= \chi_{j:j-1} \otimes \chi_{j-1:j-2} \otimes \cdots \otimes \chi_{1:0} \otimes \rho_0 \nonumber \\
    &= \bigotimes_{j=1}^{N} \chi_{j:j-1} \otimes \rho_0, \label{eq:PT-Markov-chain}
\end{align}
where $\rho_0$ is the average initial state of the process. In other words, the process is said to be a Markov chain if and only if the process tensor is a product state across time. This allows one to make use of a quasi-distance based measure of non-Markovianity:
\begin{align}
    \mathcal{N} = \underset{\xi^{\rm \small Markov}_{j:0}}{\rm min} D\big(\xi_{j:0} \Vert \xi^{\rm \small Markov}_{j:0} \big), \label{eq:PT-NM-definition}
\end{align}
which can be interpreted as the minimum distance from the closest Markov process, where $D$ could be any $CP$-contractive\footnote{Here, contractivity means that a $CP$-contractive distance must satisfy the data processing inequality under a Markov CPTP map $\Phi$: $D(\Phi[\rho] \Vert \Phi[\sigma]) \le D(\rho \Vert \sigma)$} quasi-distance such as quantum relative entropy $D(\rho \Vert \sigma) = Tr[\rho \log \rho - \rho \log \sigma]$ \cite{white2021many}. 

Here, some important remarks are in order. The definition in Eq. (\ref{eq:PT-NM-definition}) is a necessary and sufficient condition for a process to be called non-Markovian, however the converse may not be true. For example, \cite{milz2019completely} recently proposed a notion of `operational divisibility' which captures memory effect present in a CP-divisible process. Process tensor is most general enough to capture all the notions of non-Markovianity under certain limits, for example it incorporates BLP and RHP based witnesses assuming that all intermediary control operations are identity. It circumvents problems of conventional two-time maps approach when initial S-E correlations are present; it allows for completely positive, linear dynamics reconstructed from measurement data via quantum process tomography [see Ref. \cite{pollock2018non} for proofs and further details]. Process tensor also tends to a definition of classical memory when the choice of instruments as well as the causal breaks are fixed. It provides a clear operational meaning to addressing the questions of open system evolution by separating the experimenter from the underlying process that is inaccessible, making it a suitable framework to handle information processing tasks in laboratory. One such situation where one wants to remove certain unwanted memory effects arising from cross-talk was recently studied in detail by \cite{white2022filtering} using process tensor framework. 

Quantum combs, in fact, provide a unified framework to describe quantum channels with classical and quantum memory. Therefore, it is pertinent to ask how to distinguish such memory effects. Interestingly, \cite{giarmatzi2021witnessing} used the process matrix framework, proposed by \cite{oreshkov2012quantum}, to `witness' \textit{genuinely} quantum memory. It is interesting to note that the process tensor can be used to identify genuinely quantum memory effects in an arbitrary process. Considering the fact that a non-Markovian process deviates from a product of marginals given in Eq. (\ref{eq:PT-Markov-chain}), the positive value of entanglement negativity, given by $\underset{\tau_B}{\rm max} \frac{1}{2}[\Vert \xi^{\tau_B}_{j:0} \Vert_1 - 1]$, of the Choi state $\xi_{j:0}$ gives a measure of `quantumness' of non-Markovian memory, where $\tau_B$ is partial transpose over some bi-partition which could be an interval between any two time-steps \cite{white2021many}.  

\subsection{Conditional past-future correlations \label{sec:CPF-correlations}}
The notion of past-future independence as definition of Markovianity was used in \cite{li2018concepts} to analyse a hierarchy in the definition of quantum Markovianity. Recently, \cite{budini2018quantum} proposed a definition of non-Markovianity based on the violation of \textit{conditional} past-future (CPF) independence. Similar to the process tensor formalism, CPF independence is generated by the ``causal break'' in the process via intermediate measurements. Hence, it is claimed that the definition via CPF independence has operational meaning in the traditional approach \cite{budini2022non-operational}. 

In a classical stochastic process, measuring a system at three successive instances $t_a < t_b < t_c$ yields outcomes $a \rightarrow b \rightarrow c$. A Markov process gives rise to factorized joint probability conditioned on immediate past outcomes:
\begin{align}
    P(a,b,c) = P(c|b)P(b|a)P(a),
\end{align}
where P(a) is the probability that the outcome $a$ occurs, and $P(b|a)$ is the probability of $b$ occurring given that $a$ has been learned. Bayes' rule allows us to formulate the criterion for a Markovianity: That given a fixed intermediate state, the future outcomes become statistically independent from the past ones. So, we have the conditional probability of future even $c$ and past event $a$ given the present $b$ is given by
\begin{align}
    P(c,a | b ) = P(c|b) P(a|b)~. \label{eq:CPF-markov-condition}
\end{align}
This in fact can be quantified via the correlation function \cite{budini2018quantum}
\begin{align}
    C_{pf} \equiv \langle O_c O_a \rangle _b - \langle O_c \rangle _b \langle  O_a \rangle _b ~, \label{eq:CPF}
\end{align}
where the operator $O$ is specific system property one would want to measure for each system state. Given that, we can write $C_{pf}$ as
\begin{align}
    C_{pf} = \sum_{ca} [P(c,a|b) - P(c|b)P(a|b)]O_c O_a ~. \label{eq:CPF-operator-form}
\end{align}
Here, the sum is over all possible outcomes $c \in \{c_1,c_2,...\}$ and $a \in \{a_1,a_2,...\}$ that occur at $t_c \in \{t_{c_1},t_{c_1}...\} $ and $t_a \in \{t_{a_1},t_{a_1}...\} $, respectively, with a given, fixed value of $b \in \{b_1,b_2,...\}$. Generally, a quantum Markov process condition satisfies Eq. (\ref{eq:CPF-markov-condition}) in the quantum setting as well and that yields $C_{pf} = 0$, which provides a straightforward generalization of classical definition to quantum domain. 

Calculating CPF correlation measure for the quantum non-Markovian process boils down to finding the predictive and retrodictive probabilities for given system operators, and substituting them in Eq. (\ref{eq:CPF-operator-form}), which we discuss below.

Let $M_a$, $M_b$, and $M_c$ be the measurement operators successively performed on system at $t_a$, $t_b$, and $t_c$, respectively, with the condition that $\sum_j M_j^\dagger M_j = I$, where $j$=a or b or c. Given that $a$ is in the past of $b$, the conditional probability $P(a|b)$ is a retrodicted quantum probability. In terms of measurement operators $M_a$ and the past quantum state $\mathcal{\rho} \equiv (\rho_0,E_b)$, it can be written as $P(a|b) = {\rm Tr}[E_b M_a \rho_0 M_a^\dagger E_b^\dagger]$, where $\rho_0$ is the initial density matrix and $E_b = M_a^\dagger M_a$ is the `effect' operator. On the other hand, $P(c|b, a)$ is the standard predictive probability. Therefore, substituting these conditional probabilities in the LHS of Eq. (\ref{eq:CPF-markov-condition}), we have
\begin{align}
    P(c,a |b ) = {\rm Tr}(E_c \rho_b) \cdot \frac{{\rm Tr}(E_b M_a \rho_0 M_a^\dagger)}{\sum_{a'} {\rm Tr}(E_b M_{a'} \rho_0 M_{a'}^\dagger)}~.
\end{align}
Assuming that the system evolves under the action of environment, one may adopt the two-time dynamical (CPTP) map between two successive instances. Then the conditional probabilities will vary according to the intermediate evolution between measurements as:
\begin{align}
    P(c,a |b ) = {\rm Tr}(E_c \Phi^\prime[\rho_b]) \cdot \frac{{\rm Tr}(E_b \Phi[M_a \rho_0 M_a^\dagger])}{\sum_a' {\rm Tr}(E_b \Phi[M_{a'} \rho M_{a'}^\dagger)]},
\end{align}
where $\Phi = \Phi(t_b,t_a)$ and $\Phi^\prime = \Phi^\prime (t_c,t_b)$, with $\Phi_t = \exp\{t \mathcal{L}\}$ where $\mathcal{L}$ is generator of the semigroup dynamics. In general, environment interaction may generate dynamics that is not semigroup and even be non-Markovian. In that case, the dynamical map $\Phi(t,t_0)$ between two successive instances will follow the Eq. (\ref{eq:two-time-map}).  

CPF correlation measure has some interesting properties. For a non-Markovian process either $C_{pf} < 0$ or $C_{pf} > 0$. Since this criterion witnesses memory in CP-divisible processes, it may be termed a necessary and sufficient condition for non-Markovianity. And the process is Markovian iff $C_{pf} = 0$. The reader is referred to \cite{budini2018quantum} for other properties.

 \section{Quantum non-Markovianity in experiments \label{sec:experimental}}

It is now well-acknowledged that simulating open quantum systems is important for many technological applications. Memory effects could prove advantageous or disadvantageous for quantum information processing, depending on the task at hand. Therefore, it is imperative to discuss and understand aspects of simulating quantum non-Markovianity in experimental setups. In this section, we present a brief survey of experimental realizations of open system dynamics, with particular focus on experiments that also study signatures of non-Markovianity.   

One of the robust methods of simulating open system dynamics is via optical setups \cite{salles2008experimental,rossi2017non,cialdi2017all,cuevas2019all}. These setups mimic the effect of an environment on a quantum system and are seen to be effective in experimentally realizing a quantum channel. In~\cite{chiuri2012linear},  non-Markovian dynamics of a qubit attached to an ancilla and a simulated environment (an Ising chain, for instance) is experimentally implemented and the importance of strong S-E correlations in the emergence of non-Markovianity is highlighted. A similar setup is also used in~\cite{liu2018experimental}, where a simulated Ising chain in a transverse field is used as the environment to study arbitrary dephasing dynamics of a photonic qubit. Similar setups to simulate quantum channels have been used to understand the transition from Markovian to non-Markovian dynamics by controlling the S-E coupling \cite{liu2011experimental,chiuri2012linear,tang2012measuring,fisher2012optimal,lyyra2022experimental}. 

We have earlier noted that an open system (S) coupled to an ancilla (A), while undergoing non-Markovian evolution establishes  quantum correlations with A that vary non-monotonously in time. In \cite{wu2020detecting}, by coupling polarization degree (system) with the frequency degree (environment), it was experimentally demonstrated that quantum-incoherent relative entropy of coherence (QI-REC) is commensurate with the S-A entanglement, thus establishing a relationship between  QI-REC and information back-flow from environment to the system.
\begin{figure}
	\centering
	\includegraphics[width=0.62\textwidth]{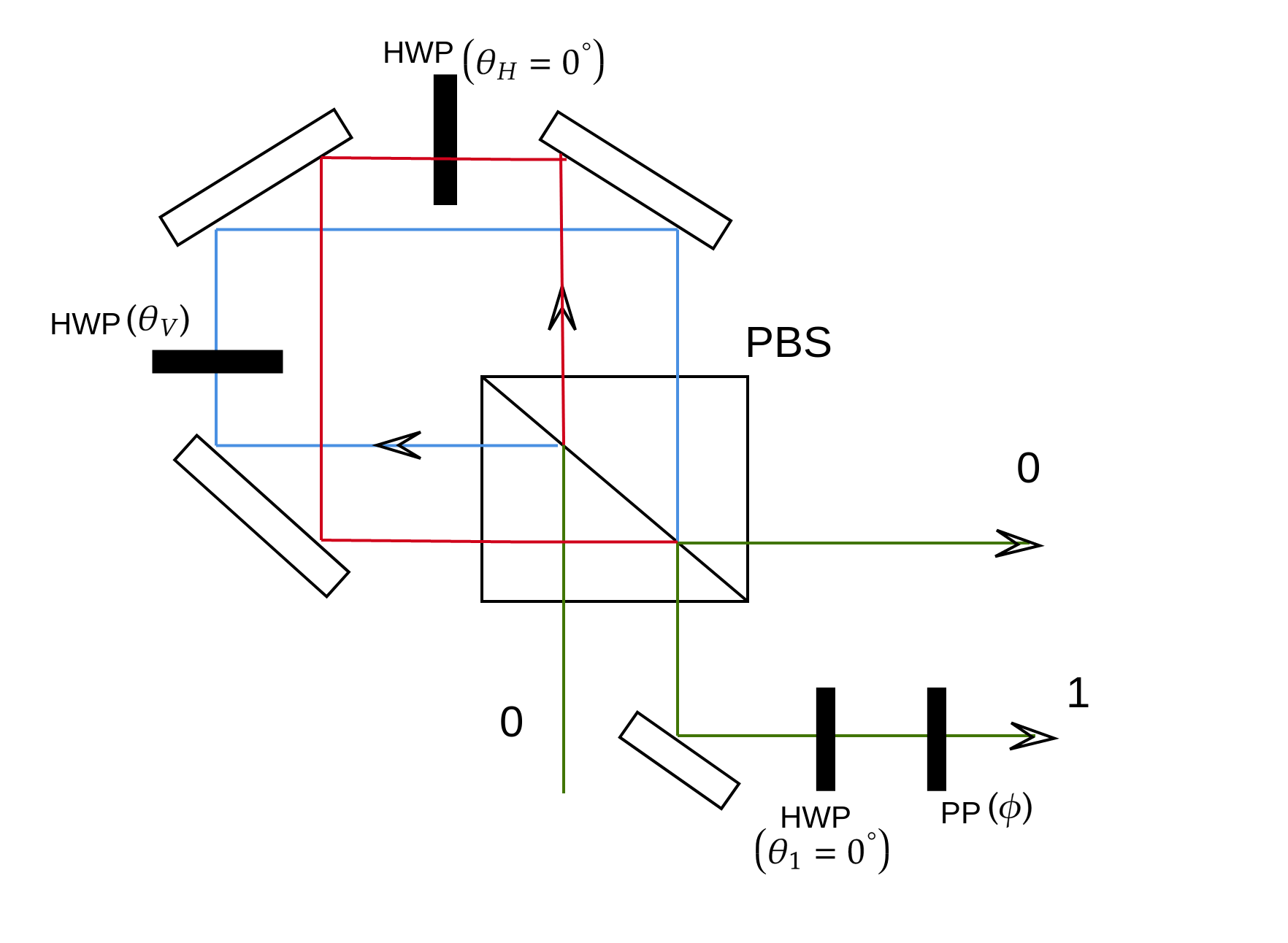}
	\caption{Optical simulation of amplitude damping channel using a Sagnac interferometer, which we have adopted from the Ref. \cite{salles2008experimental}. Here, PBS is the polarization beam splitter, HWP($\theta_V$) (the black rectangular slabs) is a Half-wave plate which rotates the vertical polarization by an angle $\theta$ and so does HWP($\theta_V$) to the horizontal polarization. PP($\phi$) is a phase plate . The white rectangular slabs are perfectly reflecting mirrors. In the above setup, setting $\phi = 0$ one realizes an amplitude damping channel for arbitrary $\theta_V$. Note that the symbols $0$ and $1$ are labels for optical modes. For different values of $\theta_H$, $\theta_V$, $\theta_1$ and $\phi$, other prototypical quantum channels can be realized [See, Table II in \cite{salles2008experimental}]. }
	\label{fig:ADsimulate}
\end{figure}

Interestingly, it is shown via experiments~\cite{farias2012observation} that when a part of a Bell pair interacts with an environment, the dynamics might lead to genuine multipartite entanglement between all the environment degrees of freedom and the initial Bell pair. When there are initial correlations in a composite environment, an open \textit{bipartite} system interacting with it might have locally Markovian evolution while globally it may show strong nonlocal memory effects \cite{laine2012nonlocal} and this counter-intuitive effect has been realized experimentally in \cite{liu2013photonic}. Quantum non-Markovianity may also arise when two Markovian channels are convex combined and \cite{uriri2020experimental} have experimentally confirmed how a convex mixture of two Pauli semigroups may result in a CP-indivisible quantum channel. In \cite{fanchini2014non-Markovianity}, the polarization degree of freedom is taken to be the two-level open system and a Sagnac interferometer is used to realize non-Markovian amplitude damping of photon polarization.

All of the above works in the optical domain use interferometric setup to simulate decoherence of quantum systems. In Fig.(\ref{fig:ADsimulate}) we give an schematic example of such a setup. Photonic simulation of quantum channels may find certain unique applications in quantum information tasks. For example, \cite{shrikant2020pingNM} showed that by deliberately adding amplitude damping quantum noise on the polarization degree of freedom (via optical simulation) in ping-pong protocol proposed by \cite{bostrom2002deterministic} one could improve security against an attack due to \cite{wojcik2003eavesdropping,bostrom2008security}. It is known that squeezing is a resource for continuous variable quantum information processing. In \cite{xiong2018optomechanical} it is shown that a cavity-optomechanical system interacting with a non-Markovian environment can lead to enhanced squeezing of mechanical mode. Therefore, exploring optical implementation of non-Markovian quantum channels may find similar counter-intuitive benefits in quantum information tasks.  

Although we have focused mainly on optical setups in this section, other platforms such as NMR \cite{bernardes2015experimental}, trapped ions \cite{wittemer2018measurement} and multi-qubit superconducting devices \cite{white2020demonstration} have also been used to implement non-Markovian open system dynamics. The importance of such experimental characterizations of non-Markovianity is further highlighted by recent works that show how unwanted memory effects that creep in due to cross-talk between superconducting qubits in quantum computer can be removed~\cite{gambetta2012characterization,white2022filtering}; however, mitigating errors due to non-Markovian memory effects arising from an uncontrollable environment can prove to be significantly harder.

\section{Afterword: Summary and Future Outlook \label{sec:conclusion}}
\begin{figure}
	\centering
	\includegraphics[width=0.85\textwidth]{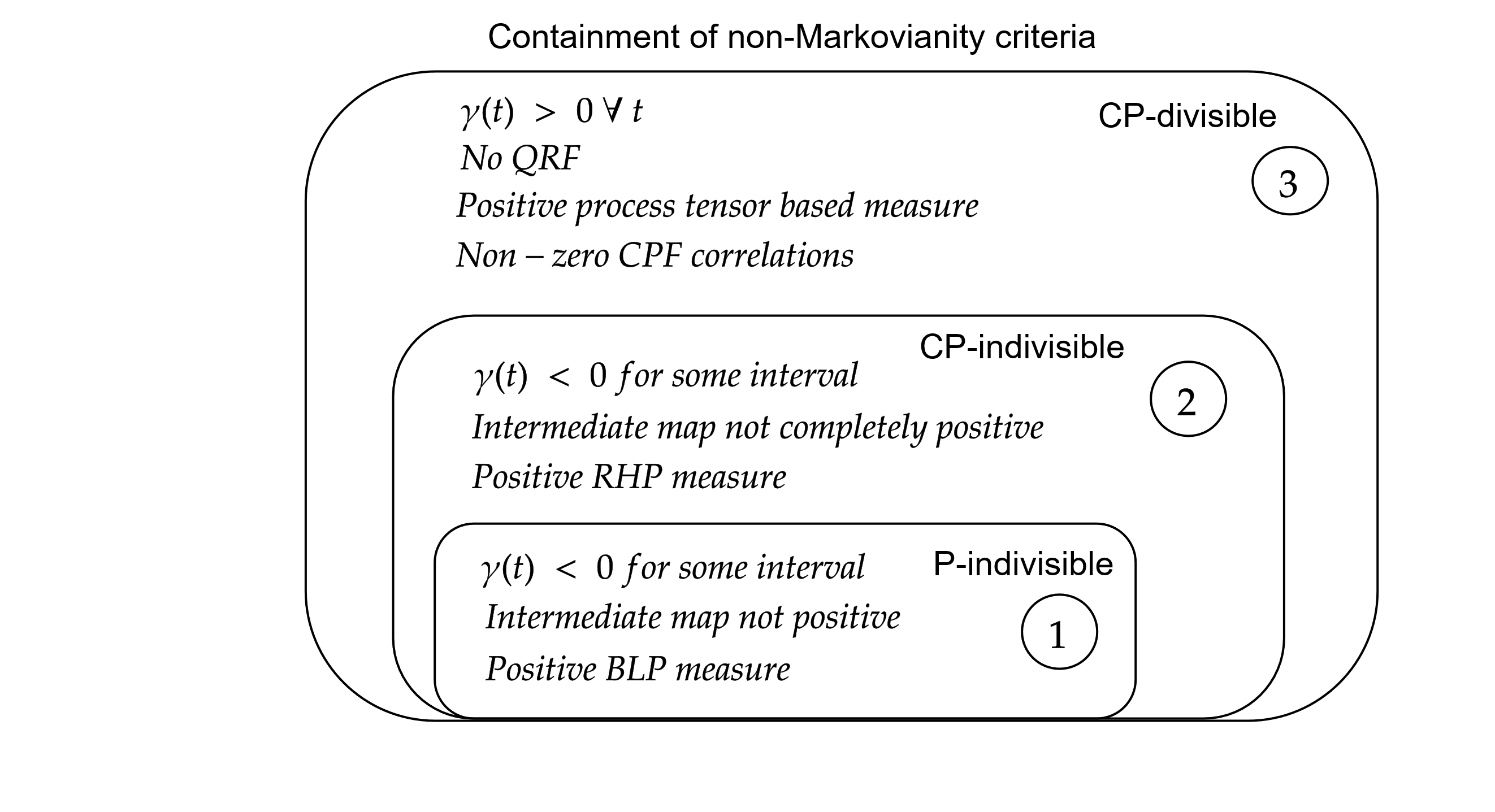}
	\caption{Containment of non-Markovianity criteria. Here, box 1 sufficiently implies 2 and 3, and 2 sufficiently implies 3, but 3 doesn't necessarily imply 2 and 1, and 2 doesn't necessarily imply 1. Here, we have depicted the containment for the processes that can be fully characterized by two-time correlations of the environment. Note that the process tensor here is for a two-time step process, and quantum regression is for two-time correlation functions. The function $\gamma(t)$ represents the decay rate in the time-local master equation.}
 \label{fig:containment}
\end{figure}

Traditionally, the dynamics of open system is described by either master equation or a two-time dynamical map \cite{breuer2002theory,banerjee2018open}. Open system evolution may be categorized mainly as Markov or non-Markov \cite{RHP14}. The precise and universal definition of non-Markovianity has remained elusive \cite{li2018concepts}, and there had been no known way of translating the classical definitions to the quantum domain until recently by \cite{pollock2018non,pollock2018operational,budini2018quantum}. We have reviewed some recent developments in the field followed by brief accounts of traditional approaches to characterizing and quantifying quantum non-Markovianity. 

Quantum causality has been a long standing puzzle within quantum theory \cite{brukner2014quantum,costa2022no,vilasini2022embedding}. It is interesting to note that quantum causality and non-Markovianity have been shown to be intimately connected via quantum temporal correlations. Notably, \cite{milz2018entanglement} showed that a causally non-separable process with a tripartite initial entangled state can simulate a multi-time non-Markovian process. In this review, we have studied how one can quantify non-Markovianity using temporal quantum correlations such as temporal steering \cite{chen2016quantifying} and causal correlations in PDM \cite{shrikant2021causal}. Note, however, that these correlations are between the states \textit{across} time. Now, it will be an interesting study to understand if PDM can offer a multi-time characterization of correlations in the process, at least for the case of qubits. In fact, \cite{zhang2020quantum} have shown that there are three different mappings from PDM to a process matrix, and since process matrix \cite{oreshkov2012quantum,costa2016quantum} and process tensor \cite{pollock2018non} essentially arise from quantum combs \cite{chiribella2009theoretical}, it would be an interesting future problem to find mappings from multi-time PDM to the process tensor, if any. 

If one studies open systems within the paradigm of two-time dynamical maps, one runs into the problem of defining a physically valid dynamical map that is both completely positive and linear when initial S-E quantum correlations are present. Pechukas's theorem states that in order to write down such a map one has to give up either complete positivity or linearity \cite{pechukas1994reduced,alicki1995comment,pechukas1995pechukas}. Later, the debate continued with regards to the nature of initial S-E correlations, for example that of quantum discord, and whether vanishing discord provided necessary and sufficient condition. However, recently some approaches have been proposed to describe dynamics of open system with initial S-E correlations in a consistent manner, some of which we have mentioned in Section \ref{sec:S-E-CPness}. 

Quantum combs \cite{chiribella2009theoretical} framework talks about the temporal correlations between observables corresponding to the dynamical process, by mapping a process to a state via CJ isomorphism. Particularly in process tensor framework \cite{pollock2018non}, temporal correlations in the multi-time description of a process corresponds to memory (or, non-Markovianity) and the framework offers incorporation of (unknown) initial S-E correlation without sacrificing linearity and complete positivity of the map. Moreover, it offers an operational definition of (non-)Markovianity via quantum process tomography, which has an appropriate classical limit. It also offers a solution to the problem of necessary and sufficient condition for non-Markovianity. 

Figure (\ref{fig:containment}) depicts the containment of non-Markovian processes according to various criteria, particularly with regards to necessary and/or sufficient condition for these criteria to witness non-Markovianity. The processes that are non-Markovian according to BLP are non-Markovian according all other criteria, hence such processes are \textit{strongly} non-Markovian. However, if BLP identifies a process as Markovian, it may still be non-Markovian according to RHP criterion hence also according to CPF correlation measure and process tensor measure. However, there may be processes that are non-Markovian according CPF correlations and process tensor, but are Markovian according to RHP. Therefore, one may conclude that Box 1 criteria are sufficient and not necessary relative to Box 2 and 3. Recent developments \cite{pollock2018operational,budini2018quantum,milz2019completely,utagi2020temporal} have shown that RHP criterion is also only sufficient but not necessary for detecting non-Markovianity relative to Box 3. Thus, one may conclude that Box 3 are necessary and sufficient criteria for quantum non-Markovianity. Interestingly, it is known that the measure based on TSW detects the range of non-Markovianity that BLP does, however it is yet to be found where the TSW measure (\ref{eq:TSmeasure2}) and the causality based measure (\ref{eq:cNMmeasure}) fit in the above containment boxes. 

From the perspective of quantum information theory, it is possible that quantum non-Markovianity may be useful in certain specific situations. In particular, since quantum non-Markovianity brings information that is ``lost" to the environment back to the system for certain time-intervals of the evolution, it might prove advantageous in certain quantum information processing tasks~\cite{bylicka2014non,laine2014nonlocal,shrikant2020pingNM}.

On the other hand, witnessing and characterizing the extent of non-Markovianity is essential in obtaining a complete understanding of noise in quantum systems. Indeed, one of the biggest challenges in scaling up quantum technologies today is to protect quantum information from environment-induced decoherence. The theory of quantum error correction (QEC) \cite{nielsen_chuang_2010} provides the means to protect information from noise, by appending a large number of physical qubits to create a single, protected logical qubit. Standard works on QEC have heavily focused on Markovian noise, and barring a couple of works~\cite{oreshkov2007, taranto2021non}, the role of QEC in mitigating noise in the non-Markovian regime remains largely unexplored. Going beyond error correction, the question of achieving quantum fault tolerance in the presence of non-Markovian quantum noise has also been explored in the past~\cite{aharonov2006fault, terhal2005fault}. Recently, there have been attempts to extend the theory of noise-adapted QEC~\cite{ng2010simple, mandayam2012towards} to non-Markovian noise models~\cite{Ng2018open-error, kwon2022reversing, lautenbacher2022}. Going forward, characterizing non-Markovianity in near-term quantum devices and developing QEC protocols adapted to non-Markovian noise promises to be an important and fruitful research avenue. 

In this contribution, we have attempted to put in perspective some of the recent developments in defining and measuring quantum non-Markovianity. It will be interesting to see how various frameworks of open system dynamics and definitions of quantum non-Markovianity allow for their uses in very specific cases of quantum information processing.

\section*{Conflict of Interest Statement}
The authors declare that the research was conducted in the absence of any commercial or financial relationships that could be construed as a potential conflict of interest.

\section*{Author Contributions}
US was the primary resource person for this article and the primary contributor. PM contributed to the writing and structural organization of this review.

\section*{Funding}
US thanks IIT Madras for the support through the Institute Postdoctoral Fellowship. This work was partially funded under Grant No. DST/ICPS/QuST/Theme-3/2019/Q59 from the Department of Science and Technology, Gov. of India and a grant from the Mphasis Foundation to the Centre for Quantum Information, Communication, and Computing (MCQuICC), IIT Madras.

\section*{Acknowledgments}
US thanks Simon Milz for insightful discussions. 

\bibliography{review}

\end{document}